\documentclass[]{beilstein}

\usepackage{tabularx}
\restylefloat{table}
\usepackage{float}
\usepackage{xspace}
\usepackage{graphicx}
\usepackage{dcolumn}
\usepackage{bm}
\usepackage[colorlinks=true,linkcolor=blue,allcolors=blue]{hyperref}
\usepackage{natbib}
\usepackage{subfigure}
\usepackage{tikz}
\usepackage{pgfplots}
\usepackage{slashbox}
\usepackage{hyperref}
\restylefloat{figure}
\nolinenumbers

\begin{document}

\title{Effect of Lubricants on the Rotational Transmission between Solid-State Gears}
\author{Huang-Hsiang Lin}
\affiliation{Institute for Materials Science and Max Bergmann Center of Biomaterials, TU Dresden,  Dresden, Germany}

\author[1]{Jonathan Heinze}

\author*[1]{Alexander Croy}{alexander.croy@tu-dresden.de}

\author[1]{Rafael Guti\'errez}

\author*[1]{Gianaurelio Cuniberti}
{gianaurelio.cuniberti@tu-dresden.de}

\maketitle

\begin{abstract}
Lubricants are widely used in macroscopic mechanical systems to reduce friction and wear. However, on the microscopic scale, it is not clear to what extent lubricants are beneficial. Therefore, in this study, we consider two diamond solid-state gears at the nanoscale immersed in different lubricant molecules and  perform classical MD simulations to investigate the rotational transmission of motion. We find that lubricants can help to synchronize the rotational transmission between gears regardless of the molecular species and the center-of-mass distance. Moreover, the influence of the angular velocity of the driving gear is investigated and shown to be  related to the bond formation process between gears.
\end{abstract}


\section{\label{sec:Introduction}Introduction}
In mechanical systems, lubrication is the most common way to reduce friction and wear\cite{Muser2006,Gnecco2006,Gnecco2015,Martini2020}. The idea of lubricants is preventing direct contact between surfaces to avoid dry friction from asperities and wear. Hence, the desirable lubrication regime would be hydrodynamic or elastohydrodynamic lubrication in the Stribeck curve\cite{Hutchings1992}. The first case corresponds to the situation that surfaces are completely separated by a fluid and the latter case is similar but surface deformations are taken into account due to high pressure at intermediate sliding velocities. On the macroscopic scale, the hydrodynamics of the fluid can be analyzed by computational fluid dynamics (CFD)\cite{Wendt2009,Blazek2015}, which is based on solving the Navier-Stokes equation\cite{ukaszewicz2016, Galdi2011} or Reynold equation\cite{Reynolds1886} for the thin-film fluid. One obtains several fluid properties such as pressure, velocity, shear stress, density and strain rate. In the case of the gear-oil-gear system, several studies based on the CFD simulation have been reported\cite{Dhar2013,Liu2018a,Liu2019a,Liu2019b,Liu2017,Guo2020,Burberi2016,Liu2018,Mastrone2020}. However, most of the simulations for this type of problem are carried out with fixed rotational speed for both gears. In this case, the gears will never be in contact with each other and only lubricant properties are calculated accordingly by the dynamical meshing at each time step. Moreover, as the system dimension approaches the nanoscale, the situation becomes very different since a continuum description of the materials might not be sufficient.

The development of the atomic force microscope (AFM)\cite{Binnig1986} and the scanning tunneling microscope (STM)\cite{Binnig1987, Binnig1982} has allowed to visualize and manipulate nanoscale gears\cite{Hla2005}. Those gears can be either solid-state gears or molecule gears, which are created by  top-down approaches (e.g.\ using focused ion beam\cite{JuYun2007} or electron beam\cite{Deng2011, Yang2014} to etch the substrate) or bottom-up approaches like chemical synthesis\cite{Gisbert2019, Abid2021}, respectively. The ultimate goal for those miniaturized gears is to implement nanoscale mechanical systems such as nanorobots\cite{Sierra2005} or mechanical calculators like the Pascaline\cite{Roegel2015}. This draws a lot of attention to issues such as triggering rotations on a surface\cite{Lin2019,PES,Croy2012,Eisenhut2018,Zhang2019,Manzano2009, Moresco2004, Moresco2015,Stolz2020, Pawin2013,Perera2013}, collective rotations\cite{Lin2019a,WeiHyo2019,Lin2020,AuYeung2020,Hove2018, Zhao2018, Hove2018a, Chen2018}, and rotational dissipation\cite{Lin2020a}. To proceed further, one may ask if lubricants can provide the same functionality as in the macroscopic case and are able to improve the transmission efficiency.

Consider the case where the lubricants film within the contact area consists only of a small number of molecules. In this case, the pressure and velocity distribution are not well defined and one has to resort to an atomistic description, for example via molecular dynamics (MD) simulations. On the other hand, the contact mechanics at the nanoscale is also very different compared to the macroscopic case since  specific pair interactions have to be taken into account by e.g.  Lennard-Jones potentials\cite{Ashcroft1976}. Several works based on MD simulations were performed to study the shear viscosity in either bulk lubricants\cite{Jadhao2017,Bair2002,Bair2002a} or lubricants confined by two surfaces\cite{Magda1985,Travis1997}. However, at the moment, MD simulations for the gear-lubricants-gear case are still missing. A deeper understanding of how different lubricants interact with gears during rotational transmission is hence highly desirable\cite{Ahmed2021}.

The paper is organized as follows: In methodology section, we introduce the setup of the gear-lubricants-gear system and details about the MD simulations. In section of results and discussions, we investigate the rotational transmission between diamond-based solid-state gears immersed in different lubricant molecules and with various center-of-mass distances. This is followed by a study of the angular velocity and the relation to the bond formation between gears.
\begin{figure*}[t]
\centering
 \includegraphics[width=\textwidth]{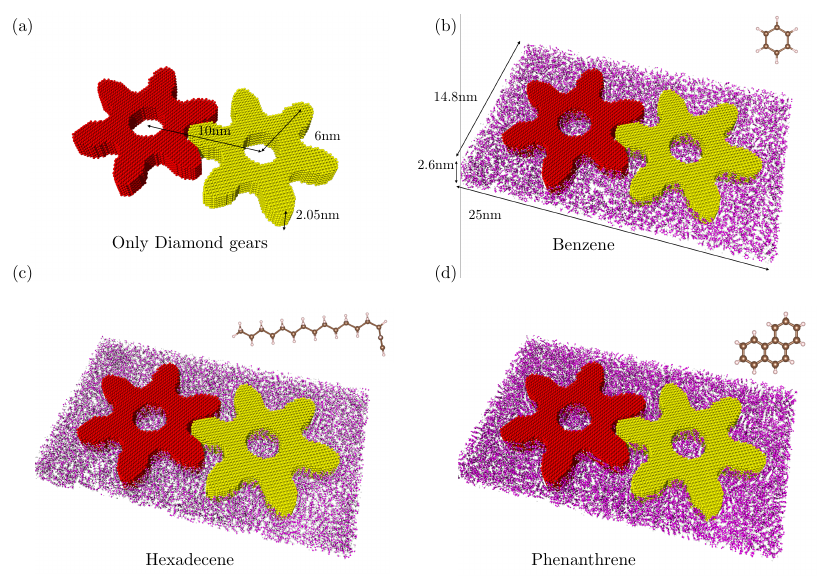}
 \caption{A schematic illustration of two interlocked diamond involute solid-state gears (colored in red and yellow) with separation distance $10$ nm, tip radius $6$ nm and thickness $2.05$ nm in the cases (a) without lubricants, lubricated by (b) benzene, (c) hexadecene and (d) phenanthrene molecules where the purple and white atoms denote the carbons and hydrogens of lubricants. The dimension of the lubricants layer shown here is $25\times 14.8 \times 2.6$ nm$^3$ for better visibility but note that the actual thickness of the lubricants layer is around $5-5.5$ nm to ensure that the gears are fully immersed.}
 \label{Fig:setup}
\end{figure*}

\section{\label{sec:Formalism} Methodologies}
In this section, we introduce how the system is defined, we specify the simulation protocols used in the MD simulations.

\subsection{\label{sec:setup}Setup}
To start our study of how different lubricants can affect the rotational transmission at the nanoscale, we consider the system shown in Fig.~\ref{Fig:setup}. First, we design two diamond solid-state gears with thickness 2.05 nm, a circular hole in the middle (with radius $r=1.5$ nm) and six involute teeth, which are optimized for transmission in classical rigid-body gears, with tip radius $r_{tip}=6$ nm. The center-of-mass distance is chosen to be 10 nm, which is large enough to ensure that there are still some lubricant molecules between the gears during rotation. To confine the gear rotation, we define an artificial Lennard-Jones (LJ) plane with parameters $\epsilon=2.875$ meV, $\sigma=3.5$ \AA{}, cutoff distance 10\AA{}, and an initial distance $9.3$ \AA{} below the gears with periodic boundary conditions in both $x$ and $y$ directions. We choose this artificial LJ plane instead of a real substrate to reduce the computational costs and to focus on the rotational transmission. 

Next, we use \textsc{gromacs}\cite{Abraham2015} to prepare the system such that the two gears are immersed in our lubricants. We have chosen  three different lubricants: benzene, 1-hexadecene and phenanthrene as shown in Fig.~\ref{Fig:setup} (b), (c) and (d), respectively. Those lubricants are chosen due to their structural simplicity and because of being liquid at room temperature. Note that in Fig.~\ref{Fig:setup} we denote the thickness of the lubricated layer as $2.6$ nm, which is only for better visibility since the actual thickness of lubricated layer is around 5-5.5 nm to ensure that the gears are immersed. Finally, the whole system is optimized by using the conjugate gradient method implemented within  \textsc{lammps}\cite{Plimpton1995}.
\begin{figure*}[t]\centering
 \includegraphics[width=1\textwidth]{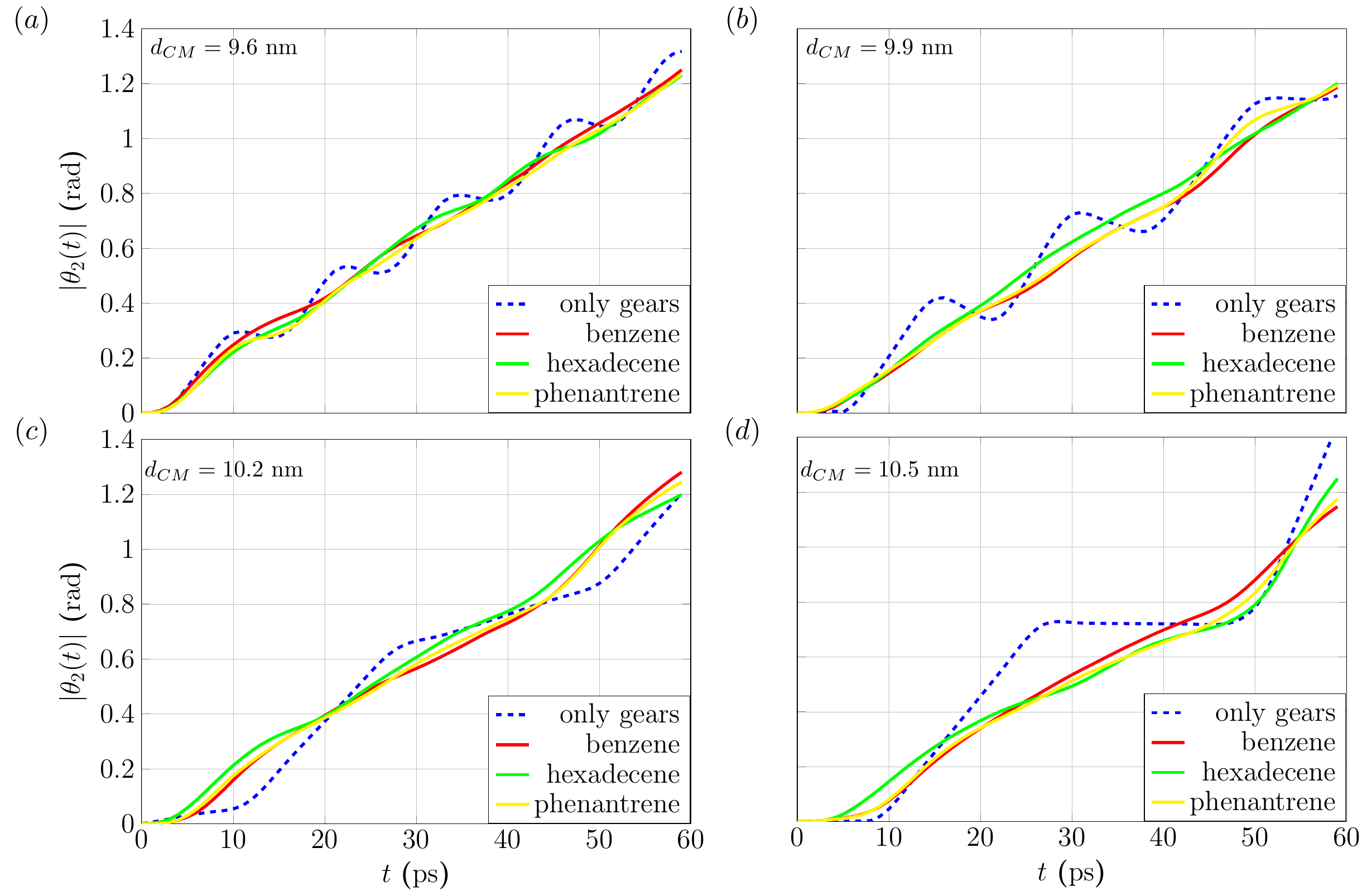}
 \caption{The trajectories of the second gear $\theta_2(t)$ in the case without (blue dashed lines) and with lubricants like benzene (red), hexadecane (green) and phenanthrene (yellow) from MD simulations within 60 ps for different center-of-mass distances $d_{CM}=$ (a) 9.6 nm,(b) 9.9 nm, (c) 10.2 nm and (d) 10.5 nm.}
 \label{Fig:transmission}
\end{figure*}

\subsection{\label{sec:MD}Molecular dynamics}
In this study, we use \textsc{lammps} to perform the MD simulation. For the force fields, we choose the Adaptive Intermolecular Reactive Empirical Bond Order (AIREBO) Potential\cite{Stuart2013} designed for hydrocarbon systems. We have used two different protocols. For protocol A, we use:
\begin{enumerate}
    \item AIREBO for carbon interactions within the gears;
    \item AIREBO for gear-lubricant-interactions;
    \item and a 12-6 Lennard-Jones potential for gear-gear-interactions.
\end{enumerate}
This protocol is only used to study the transmission between gears, since no bond formation will happen between gears. For protocol B, we use AIREBO for all interactions. In this case, we allow bonds to be formed between gears, since the AIREBO potential is reactive. On the other hand, to constrain the rotational axle, we connect a stiff spring with spring constant $k=1600$ N/m (1000 eV/\AA$^2$) to either gear's center-of-mass. To fix the temperature, the lubricants are subject to the canonical ensemble (NVT) implemented by the Nosé–Hoover thermostat\cite{Nose1984,Hoover1985} at $T=400$K and the gears are subject to the microcanonical ensemble (NVE).

\section{\label{sec:results} Results and discussions}
In this section, we present the results for the rotational transmission in the cases with and without lubricant and discuss the effect of the center-of-mass distance and of the angular velocity. Finally, we look into the bond-formation behavior between gears.

\subsection{Lubricants and distance dependence}
First, we consider the setup in Fig.~\ref{Fig:setup} with protocol A to compare how different lubricants can affect rotational transmission. We enforce a constant angular velocity $\omega=6.67\pi$ rad/ns for the first gear shown on the left in Fig.~\ref{Fig:setup}. However, instead of fixing the angular velocity for atoms in the first gear, we only fix those atoms within the inner cylindrical region, i.e., those with radius $r\leq 2$ nm. In this way, the outer atoms are allowed to deform, which avoids instantaneous torque transfer and makes the simulation more realistic. Then we monitor how the second gear can follow the motion of the first one. The results from the MD simulation within 60 ps for different center-of-mass distances, $d_{CM}=9.6$, 9.9, 10.2 and 10.5 nm, are shown in Fig.~\ref{Fig:transmission} (a), (b), (c) and (d), respectively. The blue dashed lines are trajectories of the angle of the second gear corresponding to the case without lubricants, which exhibit oscillations on top of a linearly increasing trend. One can imagine that when the angular momentum is transferred from the first gear to the second one, the teeth of both gears start to jiggling around. Moreover, there is a finite phase-shift or time delay for the second gear compared to the first one, whose trajectory is a linear straight line. This phenomenon is due to the nonzero distance between the teeth of both gears during their collective rotations. This is not expected to happen for ideal, perfectly interlocked gears in contact\cite{Han1997,Robertson1994,Lin2020}. Besides, by increasing $d_{CM}$ from 9.6 nm to 10.5 nm, one can see that this time delay behavior becomes more prominent.

For the cases with lubricants like benzene (red), hexadecene (green) and phenantrene (yellow), one can see in Fig.~\ref{Fig:transmission} that the trajectories are always smoothened and that oscillation amplitudes and time delay for the second gear are reduced. Moreover, this effect is independent of the type of lubricants and of the distance $d_{CM}$. These important findings imply that lubricants at the nanoscale can be used to synchronize both gears and make the collective rotations closer to the case of rigid bodies. The underlying reason for this behavior is the tendency of the lubricants to fill the gap between gears and to provide a medium for angular momentum transfer all time, which stabilizes the motion of the second gear. However, more energy is needed to sustain the rotation with the same angular velocity since the energy can dissipate into surrounding lubricants.
\begin{figure}[t]\centering
 \includegraphics[width=1\textwidth]{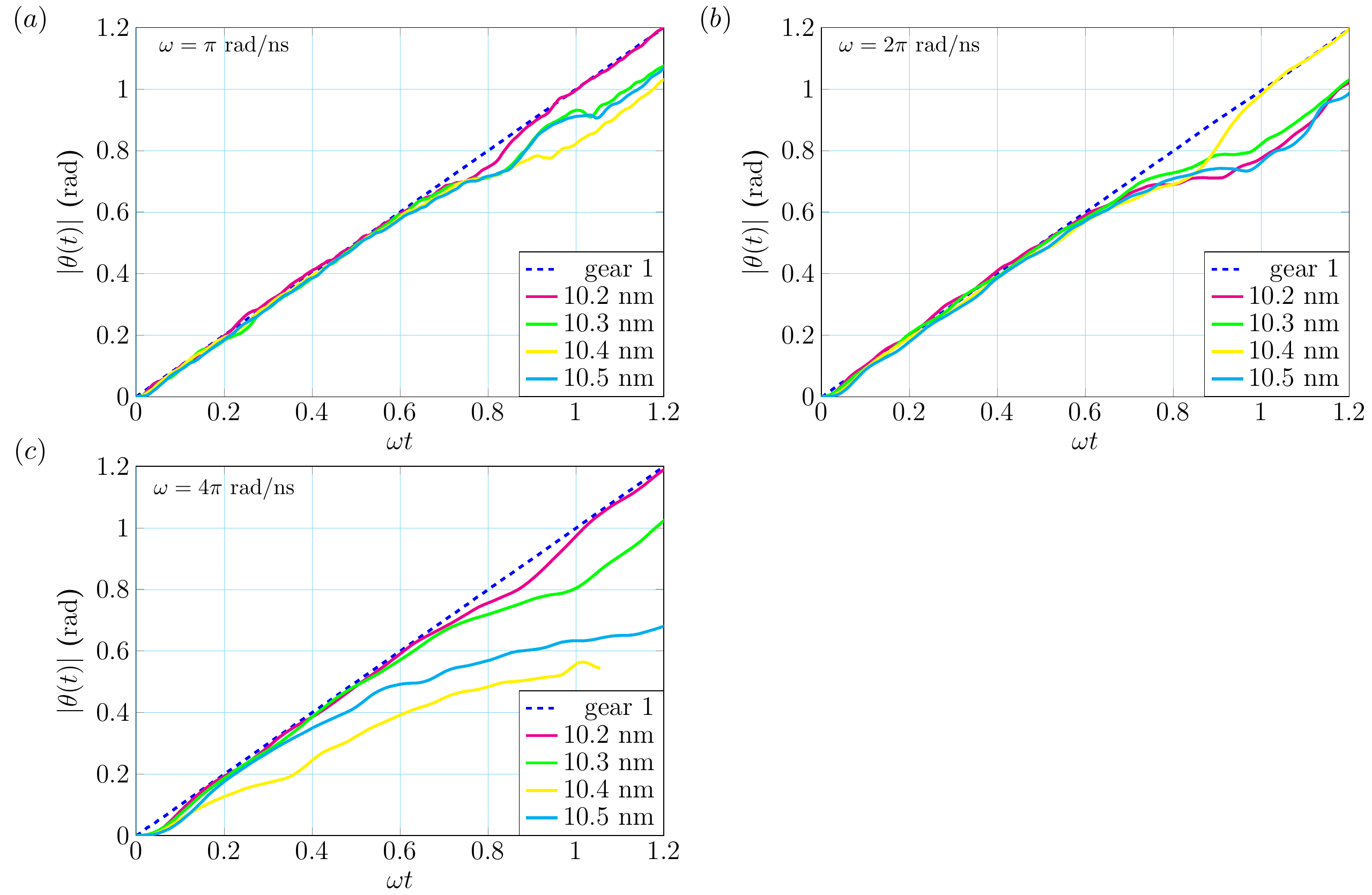}
 \caption{The trajectories from MD simulations of gears lubricated by hexadecane for different angular velocities of the first gear (a) $\omega=\pi$, (b) $\omega=2\pi$ and (c) $\omega=4\pi$ rad/ps where the blue dashed lines represent the trajectories of the first gear $\theta_1(t)$ and the others are denoting the angle $\theta_2(t)$ of the second one for different center-of-mass distances ranging from 10.2 to 10.5 nm.}
 \label{Fig:speed}
\end{figure}

\subsection{Angular velocity dependence}
From the previous section, we know that the lubricants can assist the transmission of angular momentum between gears. One might still wonder if the angular velocity of the first gear plays any role. Therefore, we performed MD simulations with different initial angular velocities $\omega=\pi$, $2\pi$ and $4\pi$ rad/ns (or periods equal to 2000, 1000 and 500 ps). To check if lubricants can protect the surface of gears,  we use  protocol B in the MD simulations, which allows the bond formation to happen between gears (C-C bond). We choose hexadecene for the following discussion. Moreover, we also run the simulations with center-of-mass distances $d_{CM}=10.2$, $10.3$, $10.4$ and $10.5$ nm to investigate how the distance dependency changes with respect to the angular velocity. Since we have different angular velocities, the simulation time is normalized in order to compare the trajectories on equal footing. We plot the trajectories with respect to the dimensionless time $\omega t$. The results are shown in Fig.~\ref{Fig:speed}. The blue dashed lines are denoting trajectories $|\theta_{1}(t)|$ from the first gear and all the other lines represent $|\theta_{2}(t)|$ for different $d_{CM}$. One can immediately see that as $\omega$ increases, some trajectories from the second gear - especially in cases of 10.4 and 10.5 nm with $\omega=4\pi$ rad/ns (yellow and cyan in Fig.~\ref{Fig:speed} (c)) - have larger phase delays. Note that for the case of 10.4 nm some atoms are lost and hence the simulation was not finished. This is due to the fact that we enforce the first gear to keep rotating after C-C bond formation between gears, which results in the dissociation of atoms from the teeth.  
\begin{figure}[t!]\centering
 \includegraphics[width=1\textwidth]{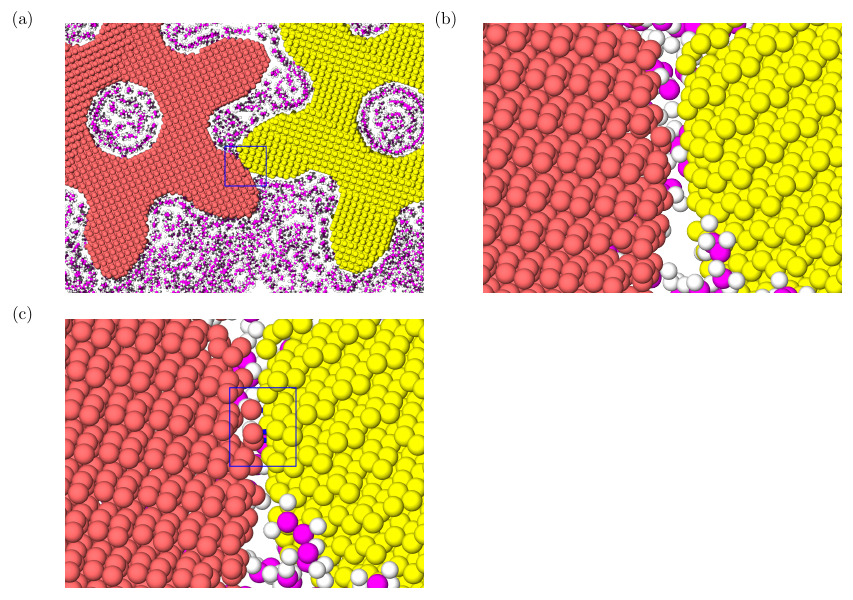}
 \caption{Snapshots for frames around the bond formation event from MD simulation with center-of-mass distance $d_{CM}=10.4$ nm and angular velocity $\omega=\pi$ rad/ns. (a) Topview of frame at $t=299$ ps ($\theta_1\approx 0.94$ rad or 54$^{\circ}$).  Zoom-in of the area around teeth (b) before  and (c) after the bond formation event ($t=300$ ps). The new bonds between gear carbon atoms (red and yellow for left and right, respectively) are marked in blue.}
 \label{Fig:bond}
\end{figure}
On the other hand, there are also some interesting phenomena happening in the cases of $d_{CM}=$10.2 nm with $\omega=\pi$ rad/ns (magenta in Fig.~\ref{Fig:speed} (a)), $d_{CM}=$10.4 nm with $\omega=2\pi$ rad/ns (yellow in Fig.~\ref{Fig:speed} (b)) and $d_{CM}=$10.2 nm with $\omega=4\pi$ rad/ns (magenta in Fig.~\ref{Fig:speed} (c)) where you can find the trajectories to have some delay around $\omega t=0.8$ and become coherent again at around $\omega t=1$. After monitoring the full trajectories, we find the time duration between $\omega t=0.8$ and $\omega t=1$ corresponds to the transition from the first step (0$^{\circ}$ to 60$^{\circ}$) to the second step (60$^{\circ}$ to 120$^{\circ}$) of the rotation. The underlying reason for the delays happening at $\omega t=0.8$ can be understood as follows (see also the movie in supplementary information): when the first step of rotation is finished, the distance between teeth is too large to effectively transfer angular momentum through the lubricants and this results in a time delay. However, when the other teeth get closer, they will then quickly repel each other again via lubricants and slightly accelerate the second gear until they become coherent again.
\begin{table}[t!]
\centering
\begin{tabular}{c|cccc}
\hline
\backslashbox{$\omega$}{$d_{CM}$ }&
\textrm{$10.2$ }&
\textrm{$10.3$ }&
\textrm{$10.4$ }&
\textrm{$10.5$ }\\
\hline
2$\pi$/3  & 50.4$^{\circ}$ & 52.0$^{\circ}$ &  61.8$^{\circ}$ & -\\
4$\pi$/5 & $-$ & $-$ & $-$ & 59.9$^{\circ}$\\
$\pi$ & $-$ &  59.0$^{\circ}$& 54.2$^{\circ}$&  59.9$^{\circ}$\\
4$\pi$/3 &$-$  &  54.0$^{\circ}$& 24.0$^{\circ}$& $-$\\
2$\pi$ & 50.8$^{\circ}$  &  54.0$^{\circ}$& $-$&  54.4$^{\circ}$\\
4$\pi$ &$-$  & 54.0$^{\circ}$ &  28.8$^{\circ}$&  38.2$^{\circ}$\\
\hline
\end{tabular}
\caption{\label{tab:table1}%
Analysis of C-C bond formation between gears for different center-of-mass distances $d_{CM}$ and angular velocities of the first gear $\omega$ in units of nm and rad/ps, respectively. The MD simulations are all performed with normalized simulation time $\omega t=1.2$ corresponding to two steps of rotation of the first gear. The values in the table represent the orientation of the first gear when the bond formation happens and a minus symbol ($-$) denotes no bond formation until the end of the simulation. 
}
\end{table}
As for the other cases, one can see that a net delay happens. Consider, for instance, $\omega=\pi$ with $d_{CM}=10.4$ nm (yellow in Fig.~\ref{Fig:speed} (a)): After investigating the full trajectory (see Fig.~\ref{Fig:bond}), we find that at the end of the transition from the first step to the second step (around $54^\circ$) some lubricants are repelled from the tooth. This process results in a net phase delay and eventually makes two gears to have direct contact with each other and to form additional C-C bonds between them (shown in Fig.~\ref{Fig:bond}~(b) and (c)). For the cases with 10.4 and 10.5 nm, these phenomena happen in the early stage of the rotations. One can see that the phase delay occurs when $\omega t\approx 0.1$ (the first gear rotates only 5$^{\circ}$) and $\omega t\approx 0.4$ (around 15$^{\circ}$) for the cases of 10.4 and 10.5 nm, respectively. This implies that for larger $d_{CM}$ and $\omega$ transmission of torque via lubricants is less effective.

\subsection{Bond formation between gears}
From the previous discussion, we know that the bond formation between gears results in a net phase delay for the second gear. Therefore, we want to investigate when the bond formation happens for different center-of-mass distances and angular velocities of the first gear. The results for different $d_{CM}$ and $\omega$ are shown in Table.~\ref{tab:table1}. Since the angular velocities are different, we use angle instead of time to compare different simulations. The values in the table show the angles for the first gear when the C-C bond formations occur between gears. One can see that, for different $d_{CM}$, there is no clear trend for the bond formations. On the other hand, we find that most bond formations happen around $50^{\circ}$ to $60^{\circ}$, which correspond to the transition from the first step to the second step since teeth are getting closer and lubricants have a higher probability to be squeezed out. As for the angular velocity, indeed we find that higher angular velocity could affect the bond formation. For instance, for $\omega=4\pi/3$ and $d_{CM}=10.4$, we have a small angle $\theta_1=24.0^{\circ}$. Moreover, for $\omega=4\pi$ and $d_{CM}=10.4$ and $10.5$, we have also two small angles $\theta_1=28.8^{\circ}$ and $38.2^{\circ}$, which means the angular velocity is very high and the lubricants cannot follow the motion of the first gear adiabatically and therefore are repelled very quickly.


\section{\label{sec:Conclusion}Conclusions and outlook}
In this study, we performed MD simulations for a system of two diamond solid-state gears with different lubricants (benzene, hexadecene and phenanthrene). We found that lubricants can be used to synchronize the collective rotations in both gears by filling the gap for better angular momentum transfer and this effect is independent of the type of the lubricant molecule. Moreover, we found that as the angular velocity of the first gear increases, the net phase delay between gears becomes prominent and this can be addressed to the C-C bond formation between gears. Further investigation of bond formation under different conditions shows that most bond formations happen during the transition from the first step to the second step. On the other hand, the center-of-mass distance does not affect the bond formation while angular velocity can indeed increase the chance since lubricants cannot follow the first gear adiabatically.

Knowing that the lubricants can help synchronize the rotational transmission, future studies will need to address if one can functionalize the gear surfaces with specific chemical groups, which can prevent the bond formation but at the same time keep gears being perfectly interlocked. On the other hand, a substrate can also provide significant friction due to electron or phonon excitations\cite{Volokitin}, which cannot be captured by a Lennard-Jones plane used in our simulations. To further investigate those open questions, a more powerful pair potential such as the reactive force field (ReaxFF)\cite{Chenoweth2008} or a deep learning force field\cite{Wang2018} approach might be suitable to address the problem. Finally, we hope that full atom simulations in combination with further advances in the fabrication and chemical synthesis techniques will improve the design of solid-state gears at the microscopic scale.

\begin{acknowledgements}
This work has been supported by the International Max Planck Research School (IMPRS) for ``Many-Particle Systems in Structured Environments'' and also by the European Union Horizon 2020 FET Open project ``Mechanics with Molecules'' (MEMO, grant nr.\ 766864). We also acknowledge the Center for Information Services and High Performance Computing (ZIH) at TU Dresden for computational resources.
\end{acknowledgements}



\bibliography{paper}

\begin{thebibliography}{67}

\setboolean{nobreakdashused}{false}\bibitem[M{\"{u}}ser(2006)]{Muser2006}
M{\"{u}}ser,~M.~H. \emph{Lect. Notes Phys.};
\newblock 2006;
\newblock Vol. 704, pp 65\mynobreakdash \mynobreakdash 104. \url{doi:
  10.1007/3-540-35284-8_4}.


\setboolean{nobreakdashused}{false}\bibitem[Gnecco and Mayer(2007)]{Gnecco2006}
Gnecco,~E.; Mayer,~E. \emph{{Fundamentals of Friction and Wear}};
\newblock NanoScience and Technology;
\newblock Springer Berlin Heidelberg: Berlin, Heidelberg, 2007;
\newblock \url{doi: 10.1007/978-3-540-36807-6}.


\setboolean{nobreakdashused}{false}\bibitem[Gnecco and Meyer(2015)]{Gnecco2015}
Gnecco,~E.; Meyer,~E. \emph{Elem. Frict. Theory Nanotribology};
\newblock Cambridge University Press: Cambridge, 2015;
\newblock pp 1\mynobreakdash \mynobreakdash 303. \url{doi:
  10.1017/CBO9780511795039}.


\setboolean{nobreakdashused}{false}\bibitem[Martini et~al.(2020)Martini, Eder,
  and D{\"{o}}rr]{Martini2020}
Martini,~A.; Eder,~S.~J.; D{\"{o}}rr,~N. \emph{Lubricants} \textbf{2020},
  \emph{8} (4), 1\mynobreakdash \mynobreakdash 20. \url{doi:
  10.3390/LUBRICANTS8040044}.


\setboolean{nobreakdashused}{false}\bibitem[Hutchings(1992)]{Hutchings1992}
Hutchings,~I. \emph{Mater. Des.} \textbf{1992}, \emph{13} (3), 187. \url{doi:
  10.1016/0261-3069(92)90241-9}.


\setboolean{nobreakdashused}{false}\bibitem[Wendt(2009)]{Wendt2009}
\emph{{Computational Fluid Dynamics}};
\newblock Wendt,~J.~F., Ed.;
\newblock Springer Berlin Heidelberg: Berlin, Heidelberg, 2009;
\newblock \url{doi: 10.1007/978-3-540-85056-4}.


\setboolean{nobreakdashused}{false}\bibitem[Blazek(2015)]{Blazek2015}
Blazek,~J. \emph{{Computational Fluid Dynamics: Principles and Applications}},
  3rd ed.;
\newblock Elsevier: Oxford, 2015;
\newblock \url{doi: 10.1016/C2013-0-19038-1}.


\setboolean{nobreakdashused}{false}\bibitem[{\L}ukaszewicz and
  Kalita(2016)]{ukaszewicz2016}
{\L}ukaszewicz,~G.; Kalita,~P. \emph{{Navier–Stokes Equations}};
\newblock Advances in Mechanics and Mathematics;
\newblock Springer International Publishing: Cham, 2016;
\newblock \url{doi: 10.1007/978-3-319-27760-8}.


\setboolean{nobreakdashused}{false}\bibitem[Galdi(2011)]{Galdi2011}
Galdi,~G. \emph{{An Introduction to the Mathematical Theory of the
  Navier-Stokes Equations}};
\newblock Springer Monographs in Mathematics;
\newblock Springer New York: New York, NY, 2011;
\newblock \url{doi: 10.1007/978-0-387-09620-9}.


\setboolean{nobreakdashused}{false}\bibitem[Reynolds(1886)]{Reynolds1886}
Reynolds,~O. \emph{Philos. Trans. R. Soc. London} \textbf{1886}, \emph{177},
  157\mynobreakdash \mynobreakdash 234. \url{doi: 10.1098/rstl.1886.0005}.


\setboolean{nobreakdashused}{false}\bibitem[Dhar and Vacca(2013)]{Dhar2013}
Dhar,~S.; Vacca,~A. \emph{Tribol. Int.} \textbf{2013}, \emph{62},
  78\mynobreakdash \mynobreakdash 90. \url{doi:
  10.1016/j.triboint.2013.02.008}.


\setboolean{nobreakdashused}{false}\bibitem[Liu et~al.(2019)Liu, Link, Lohner,
  and Stahl]{Liu2018a}
Liu,~H.; Link,~F.; Lohner,~T.; Stahl,~K. \emph{Proc. Inst. Mech. Eng. Part C J.
  Mech. Eng. Sci.} \textbf{2019}, \emph{233} (21-22), 7412\mynobreakdash
  \mynobreakdash 7422. \url{doi: 10.1177/0954406219865920}.


\setboolean{nobreakdashused}{false}\bibitem[Liu et~al.(2019)Liu, Arfaoui,
  Stanic, Montigny, Jurkschat, Lohner, and Stahl]{Liu2019a}
Liu,~H.; Arfaoui,~G.; Stanic,~M.; Montigny,~L.; Jurkschat,~T.; Lohner,~T.;
  Stahl,~K. \emph{Proc. Inst. Mech. Eng. Part J J. Eng. Tribol.} \textbf{2019},
  \emph{233} (1), 74\mynobreakdash \mynobreakdash 86. \url{doi:
  10.1177/1350650118760626}.


\setboolean{nobreakdashused}{false}\bibitem[Liu et~al.(2019)Liu, Link, Lohner,
  and Stahl]{Liu2019b}
Liu,~H.; Link,~F.; Lohner,~T.; Stahl,~K. \emph{Proc. Inst. Mech. Eng. Part C J.
  Mech. Eng. Sci.} \textbf{2019}, \emph{233} (21-22), 7412\mynobreakdash
  \mynobreakdash 7422. \url{doi: 10.1177/0954406219865920}.


\setboolean{nobreakdashused}{false}\bibitem[Liu et~al.(2017)Liu, Jurkschat,
  Lohner, and Stahl]{Liu2017}
Liu,~H.; Jurkschat,~T.; Lohner,~T.; Stahl,~K. \emph{Tribol. Int.}
  \textbf{2017}, \emph{109}, 346\mynobreakdash \mynobreakdash 354. \url{doi:
  10.1016/j.triboint.2016.12.042}.


\setboolean{nobreakdashused}{false}\bibitem[Guo et~al.(2020)Guo, Chen, Liu,
  Wang, and Wang]{Guo2020}
Guo,~D.; Chen,~F.; Liu,~J.; Wang,~Y.; Wang,~X. \emph{Tribol. Trans.}
  \textbf{2020}, \emph{63} (1), 182\mynobreakdash \mynobreakdash 193. \url{doi:
  10.1080/10402004.2019.1682212}.


\setboolean{nobreakdashused}{false}\bibitem[Burberi et~al.(2016)Burberi,
  Fondelli, Andreini, Facchini, and Cipolla]{Burberi2016}
Burberi,~E.; Fondelli,~T.; Andreini,~A.; Facchini,~B.; Cipolla,~L. \emph{Proc.
  ASME Turbo Expo} \textbf{2016}, \emph{5A-2016}, 1\mynobreakdash
  \mynobreakdash 12. \url{doi: 10.1115/GT2016-57454}.


\setboolean{nobreakdashused}{false}\bibitem[Liu et~al.(2018)Liu, Jurkschat,
  Lohner, and Stahl]{Liu2018}
Liu,~H.; Jurkschat,~T.; Lohner,~T.; Stahl,~K. \emph{Lubricants} \textbf{2018},
  \emph{6} (2), 47. \url{doi: 10.3390/lubricants6020047}.


\setboolean{nobreakdashused}{false}\bibitem[Mastrone et~al.(2020)Mastrone,
  Hartono, Chernoray, and Concli]{Mastrone2020}
Mastrone,~M.~N.; Hartono,~E.~A.; Chernoray,~V.; Concli,~F. \emph{Tribol. Int.}
  \textbf{2020}, \emph{151} (June), 106496. \url{doi:
  10.1016/j.triboint.2020.106496}.


\setboolean{nobreakdashused}{false}\bibitem[Binnig et~al.(1986)Binnig, Quate,
  and Gerber]{Binnig1986}
Binnig,~G.; Quate,~C.~F.; Gerber,~C. \emph{Phys. Rev. Lett.} \textbf{1986},
  \emph{56} (9), 930\mynobreakdash \mynobreakdash 933. \url{doi:
  10.1103/PhysRevLett.56.930}.


\setboolean{nobreakdashused}{false}\bibitem[Binnig and
  Rohrer(1987)]{Binnig1987}
Binnig,~G.; Rohrer,~H. \emph{Rev. Mod. Phys.} \textbf{1987}, \emph{59} (3),
  615\mynobreakdash \mynobreakdash 625. \url{doi: 10.1103/RevModPhys.59.615}.


\setboolean{nobreakdashused}{false}\bibitem[Binnig et~al.(1982)Binnig, Rohrer,
  Gerber, and Weibel]{Binnig1982}
Binnig,~G.; Rohrer,~H.; Gerber,~C.; Weibel,~E. \emph{Appl. Phys. Lett.}
  \textbf{1982}, \emph{40} (2), 178\mynobreakdash \mynobreakdash 180. \url{doi:
  10.1063/1.92999}.


\setboolean{nobreakdashused}{false}\bibitem[Hla(2005)]{Hla2005}
Hla,~S.-W. \emph{J. Vac. Sci. Technol. B Microelectron. Nanom. Struct.}
  \textbf{2005}, \emph{23} (4), 1351. \url{doi: 10.1116/1.1990161}.


\setboolean{nobreakdashused}{false}\bibitem[{Ju Yun} et~al.(2007){Ju Yun},
  {Seong Ah}, Kim, {Soo Yun}, {Chon Park}, and {Han Ha}]{JuYun2007}
{Ju Yun},~Y.; {Seong Ah},~C.; Kim,~S.; {Soo Yun},~W.; {Chon Park},~B.; {Han
  Ha},~D. \emph{Nanotechnology} \textbf{2007}, \emph{18} (50), 505304.
  \url{doi: 10.1088/0957-4484/18/50/505304}.


\setboolean{nobreakdashused}{false}\bibitem[Deng et~al.(2011)Deng, Troadec,
  Ample, and Joachim]{Deng2011}
Deng,~J.; Troadec,~C.; Ample,~F.; Joachim,~C. \emph{Nanotechnology}
  \textbf{2011}, \emph{22} (27), 275307. \url{doi:
  10.1088/0957-4484/22/27/275307}.


\setboolean{nobreakdashused}{false}\bibitem[Yang et~al.(2014)Yang, Deng,
  Troadec, Ondar{\c{c}}uhu, and Joachim]{Yang2014}
Yang,~J.; Deng,~J.; Troadec,~C.; Ondar{\c{c}}uhu,~T.; Joachim,~C.
  \emph{Nanotechnology} \textbf{2014}, \emph{25} (46), 465305. \url{doi:
  10.1088/0957-4484/25/46/465305}.


\setboolean{nobreakdashused}{false}\bibitem[Gisbert et~al.(2019)Gisbert, Abid,
  Bertrand, Saffon-Merceron, Kammerer, and Rapenne]{Gisbert2019}
Gisbert,~Y.; Abid,~S.; Bertrand,~G.; Saffon-Merceron,~N.; Kammerer,~C.;
  Rapenne,~G. \emph{Chem. Commun.} \textbf{2019}, \emph{55} (97),
  14689\mynobreakdash \mynobreakdash 14692. \url{doi: 10.1039/C9CC08384G}.


\setboolean{nobreakdashused}{false}\bibitem[Abid et~al.(2021)Abid, Gisbert,
  Kojima, Saffon-Merceron, Cuny, Kammerer, and Rapenne]{Abid2021}
Abid,~S.; Gisbert,~Y.; Kojima,~M.; Saffon-Merceron,~N.; Cuny,~J.; Kammerer,~C.;
  Rapenne,~G. \emph{Chem. Sci.} \textbf{2021}, \emph{12} (13),
  4709\mynobreakdash \mynobreakdash 4721. \url{doi: 10.1039/D0SC06379G}.


\setboolean{nobreakdashused}{false}\bibitem[Sierra et~al.(2005)Sierra, Weir,
  Jones, and Frank]{Sierra2005}
Sierra,~D.~P.; Weir,~N.~A.; Jones,~J.; Frank, \emph{Sandia Report}
  \textbf{2005}. \url{doi: 10.2172/875622}.


\setboolean{nobreakdashused}{false}\bibitem[Roegel(2015)]{Roegel2015}
Roegel,~D. \emph{IEEE Ann. Hist. Comput.} \textbf{2015}, \emph{37} (4),
  90\mynobreakdash \mynobreakdash 96. \url{doi: 10.1109/MAHC.2015.79}.


\setboolean{nobreakdashused}{false}\bibitem[Lin et~al.(2019)Lin, Croy,
  Gutierrez, Joachim, and Cuniberti]{Lin2019}
Lin,~H.~H.; Croy,~A.; Gutierrez,~R.; Joachim,~C.; Cuniberti,~G. \emph{J. Phys.
  Commun.} \textbf{2019}, \emph{3} (2), 025011. \url{doi:
  10.1088/2399-6528/ab0731}.


\setboolean{nobreakdashused}{false}\bibitem[Echeverria et~al.(2014)Echeverria,
  Monturet, and Joachim]{PES}
Echeverria,~J.; Monturet,~S.; Joachim,~C. \emph{Nanoscale} \textbf{2014},
  \emph{6} (5), 2793. \url{doi: 10.1039/c3nr05814j}.


\setboolean{nobreakdashused}{false}\bibitem[Croy and Eisfeld(2012)]{Croy2012}
Croy,~A.; Eisfeld,~A. \emph{EPL (Europhysics Lett.} \textbf{2012}, \emph{98}
  (6), 68004. \url{doi: 10.1209/0295-5075/98/68004}.


\setboolean{nobreakdashused}{false}\bibitem[Eisenhut et~al.(2018)Eisenhut,
  Meyer, Kr{\"{u}}ger, Ohmann, Cuniberti, and Moresco]{Eisenhut2018}
Eisenhut,~F.; Meyer,~J.; Kr{\"{u}}ger,~J.; Ohmann,~R.; Cuniberti,~G.;
  Moresco,~F. \emph{Surf. Sci.} \textbf{2018}, \emph{678} (May),
  177\mynobreakdash \mynobreakdash 182. \url{doi: 10.1016/j.susc.2018.05.003}.


\setboolean{nobreakdashused}{false}\bibitem[Zhang et~al.(2019)Zhang, Calupitan,
  Rojas, Tumbleson, Erbland, Kammerer, Ajayi, Wang, Curtiss, Ngo, Ulloa,
  Rapenne, and Hla]{Zhang2019}
Zhang,~Y.; Calupitan,~J.~P.; Rojas,~T.; Tumbleson,~R.; Erbland,~G.;
  Kammerer,~C.; Ajayi,~T.~M.; Wang,~S.; Curtiss,~L.~A.; Ngo,~A.~T.;
  Ulloa,~S.~E.; Rapenne,~G.; Hla,~S.~W. \emph{Nat. Commun.} \textbf{2019},
  \emph{10} (1), 3742. \url{doi: 10.1038/s41467-019-11737-1}.


\setboolean{nobreakdashused}{false}\bibitem[Manzano et~al.(2009)Manzano, Soe,
  Wong, Ample, Gourdon, Chandrasekhar, and Joachim]{Manzano2009}
Manzano,~C.; Soe,~W.-H.; Wong,~H.~S.; Ample,~F.; Gourdon,~A.;
  Chandrasekhar,~N.; Joachim,~C. \emph{Nat. Mater.} \textbf{2009}, \emph{8}
  (7), 576\mynobreakdash \mynobreakdash 579. \url{doi: 10.1038/nmat2467}.


\setboolean{nobreakdashused}{false}\bibitem[Moresco(2004)]{Moresco2004}
Moresco,~F. \emph{Phys. Rep.} \textbf{2004}, \emph{399} (4), 175\mynobreakdash
  \mynobreakdash 225. \url{doi: 10.1016/j.physrep.2004.08.001}.


\setboolean{nobreakdashused}{false}\bibitem[Moresco(2015)]{Moresco2015}
Moresco,~F. {Driving Molecular Machines Using the Tip of a Scanning Tunneling
  Microscope}. In \emph{Single Mol. Mach. Mot.}; Joachim,~C., Rapenne,~G.,
  Eds.;
\newblock Springer International Publishing: Cham, 2015;
\newblock pp 165\mynobreakdash \mynobreakdash 186. \url{doi:
  10.1007/978-3-319-13872-5_10}.


\setboolean{nobreakdashused}{false}\bibitem[Stolz et~al.(2020)Stolz,
  Gr{\"{o}}ning, Prinz, Brune, and Widmer]{Stolz2020}
Stolz,~S.; Gr{\"{o}}ning,~O.; Prinz,~J.; Brune,~H.; Widmer,~R. \emph{Proc.
  Natl. Acad. Sci.} \textbf{2020}, \emph{117} (26), 14838\mynobreakdash
  \mynobreakdash 14842. \url{doi: 10.1073/pnas.1918654117}.


\setboolean{nobreakdashused}{false}\bibitem[Pawin et~al.(2013)Pawin, Stieg,
  Skibo, Grisolia, Schilittler, Langlais, Tateyama, Joachim, and
  Gimzewski]{Pawin2013}
Pawin,~G.; Stieg,~A.~Z.; Skibo,~C.; Grisolia,~M.; Schilittler,~R.~R.;
  Langlais,~V.; Tateyama,~Y.; Joachim,~C.; Gimzewski,~J.~K. \emph{Langmuir}
  \textbf{2013}, \emph{29} (24), 7309\mynobreakdash \mynobreakdash 7317.
  \url{doi: 10.1021/la304634n}.


\setboolean{nobreakdashused}{false}\bibitem[Perera et~al.(2013)Perera, Ample,
  Kersell, Zhang, Vives, Echeverria, Grisolia, Rapenne, Joachim, and
  Hla]{Perera2013}
Perera,~U. G.~E.; Ample,~F.; Kersell,~H.; Zhang,~Y.; Vives,~G.; Echeverria,~J.;
  Grisolia,~M.; Rapenne,~G.; Joachim,~C.; Hla,~S.-W. \emph{Nat. Nanotechnol.}
  \textbf{2013}, \emph{8} (1), 46\mynobreakdash \mynobreakdash 51. \url{doi:
  10.1038/nnano.2012.218}.


\setboolean{nobreakdashused}{false}\bibitem[Lin et~al.(2020)Lin, Croy,
  Gutierrez, Joachim, and Cuniberti]{Lin2019a}
Lin,~H.-H.; Croy,~A.; Gutierrez,~R.; Joachim,~C.; Cuniberti,~G. \emph{Phys.
  Rev. Appl.} \textbf{2020}, \emph{13} (3), 034024. \url{doi:
  10.1103/PhysRevApplied.13.034024}.


\setboolean{nobreakdashused}{false}\bibitem[Soe et~al.(2019)Soe, Srivastava,
  and Joachim]{WeiHyo2019}
Soe,~W.-H.; Srivastava,~S.; Joachim,~C. \emph{J. Phys. Chem. Lett.}
  \textbf{2019}, \emph{10} (21), 6462\mynobreakdash \mynobreakdash 6467.
  \url{doi: 10.1021/acs.jpclett.9b02259}.


\setboolean{nobreakdashused}{false}\bibitem[Lin et~al.(2020)Lin, Heinze, Croy,
  Gutierrez, and Cuniberti]{Lin2020}
Lin,~H.-H.; Heinze,~J.; Croy,~A.; Gutierrez,~R.; Cuniberti,~G. {Mechanical
  Transmission of Rotation for Molecule Gears and Solid-State Gears}. In
  \emph{Build. Probing Small Mech.};
\newblock Springer, Cham, 2020;
\newblock Chapter~11, pp 165\mynobreakdash \mynobreakdash 180. \url{doi:
  10.1007/978-3-030-56777-4_11}.


\setboolean{nobreakdashused}{false}\bibitem[{Au Yeung} et~al.(2020){Au Yeung},
  K{\"{u}}hne, Eisenhut, Kleinw{\"{a}}chter, Gisbert, Robles, Lorente,
  Cuniberti, Joachim, Rapenne, Kammerer, and Moresco]{AuYeung2020}
{Au Yeung},~K.~H.; K{\"{u}}hne,~T.; Eisenhut,~F.; Kleinw{\"{a}}chter,~M.;
  Gisbert,~Y.; Robles,~R.; Lorente,~N.; Cuniberti,~G.; Joachim,~C.;
  Rapenne,~G.; Kammerer,~C.; Moresco,~F. \emph{J. Phys. Chem. Lett.}
  \textbf{2020}, No. 111,  6892\mynobreakdash \mynobreakdash 6899. \url{doi:
  10.1021/acs.jpclett.0c01747}.


\setboolean{nobreakdashused}{false}\bibitem[Zhao et~al.(2018)Zhao, Qi, Zhao,
  Hermann, Zhang, and {Van Hove}]{Hove2018}
Zhao,~R.; Qi,~F.; Zhao,~Y.-L.; Hermann,~K.~E.; Zhang,~R.-Q.; {Van Hove},~M.~A.
  \emph{J. Phys. Chem. Lett.} \textbf{2018}, \emph{9} (10), 2611\mynobreakdash
  \mynobreakdash 2619. \url{doi: 10.1021/acs.jpclett.8b00676}.


\setboolean{nobreakdashused}{false}\bibitem[Zhao et~al.(2018)Zhao, Qi, Zhang,
  and {Van Hove}]{Zhao2018}
Zhao,~R.; Qi,~F.; Zhang,~R.-q.; {Van Hove},~M.~A. \emph{J. Phys. Chem. C}
  \textbf{2018}, \emph{122} (43), 25067\mynobreakdash \mynobreakdash 25074.
  \url{doi: 10.1021/acs.jpcc.8b08158}.


\setboolean{nobreakdashused}{false}\bibitem[Zhao et~al.(2018)Zhao, Zhao, Qi,
  Hermann, Zhang, and {Van Hove}]{Hove2018a}
Zhao,~R.; Zhao,~Y.-L.; Qi,~F.; Hermann,~K.~E.; Zhang,~R.-Q.; {Van Hove},~M.~A.
  \emph{ACS Nano} \textbf{2018}, \emph{12} (3), 3020\mynobreakdash
  \mynobreakdash 3029. \url{doi: 10.1021/acsnano.8b00784}.


\setboolean{nobreakdashused}{false}\bibitem[Chen et~al.(2018)Chen, Qi,
  Jitapunkul, Zhao, Zhang, and {Van Hove}]{Chen2018}
Chen,~L.; Qi,~F.; Jitapunkul,~K.; Zhao,~Y.; Zhang,~R.; {Van Hove},~M.~A.
  \emph{J. Phys. Chem. A} \textbf{2018}, \emph{122} (38), 7614\mynobreakdash
  \mynobreakdash 7619. \url{doi: 10.1021/acs.jpca.8b04368}.


\setboolean{nobreakdashused}{false}\bibitem[Lin et~al.(2021)Lin, Croy,
  Gutierrez, and Cuniberti]{Lin2020a}
Lin,~H.~H.; Croy,~A.; Gutierrez,~R.; Cuniberti,~G. \emph{Phys. Rev. Appl.}
  \textbf{2021}, \emph{15} (2), 024053. \url{doi:
  10.1103/PhysRevApplied.15.024053}.


\setboolean{nobreakdashused}{false}\bibitem[Ashcroft and
  Mermin(1976)]{Ashcroft1976}
Ashcroft,~N.~W.; Mermin,~N.~D. \emph{{Solid State Physics}};
\newblock Holt, Rinehart and Winston, 1976.


\setboolean{nobreakdashused}{false}\bibitem[Jadhao and
  Robbins(2017)]{Jadhao2017}
Jadhao,~V.; Robbins,~M.~O. \emph{Proc. Natl. Acad. Sci.} \textbf{2017},
  \emph{114} (30), 7952\mynobreakdash \mynobreakdash 7957. \url{doi:
  10.1073/pnas.1705978114}.


\setboolean{nobreakdashused}{false}\bibitem[Bair et~al.(2002)Bair, McCabe, and
  Cummings]{Bair2002}
Bair,~S.; McCabe,~C.; Cummings,~P.~T. \emph{Phys. Rev. Lett.} \textbf{2002},
  \emph{88} (5), 058302. \url{doi: 10.1103/PhysRevLett.88.058302}.


\setboolean{nobreakdashused}{false}\bibitem[Bair et~al.(2002)Bair, McCabe, and
  Cummings]{Bair2002a}
Bair,~S.; McCabe,~C.; Cummings,~P.~T. \emph{Tribol. Lett.} \textbf{2002},
  \emph{13} (4), 251\mynobreakdash \mynobreakdash 254. \url{doi:
  10.1023/A:1021011225316}.


\setboolean{nobreakdashused}{false}\bibitem[Magda et~al.(1985)Magda, Tirrell,
  and Davis]{Magda1985}
Magda,~J.~J.; Tirrell,~M.; Davis,~H.~T. \emph{J. Chem. Phys.} \textbf{1985},
  \emph{83} (4), 1888\mynobreakdash \mynobreakdash 1901. \url{doi:
  10.1063/1.449375}.


\setboolean{nobreakdashused}{false}\bibitem[Travis et~al.(1997)Travis, Todd,
  and Evans]{Travis1997}
Travis,~K.~P.; Todd,~B.~D.; Evans,~D.~J. \emph{Phys. Rev. E} \textbf{1997},
  \emph{55} (4), 4288\mynobreakdash \mynobreakdash 4295. \url{doi:
  10.1103/PhysRevE.55.4288}.


\setboolean{nobreakdashused}{false}\bibitem[Ahmed et~al.(2021)Ahmed, Ullah,
  Zhao, Zhang, and {Van Hove}]{Ahmed2021}
Ahmed,~S.~B.; Ullah,~N.; Zhao,~Y.; Zhang,~R.; {Van Hove},~M.~A. \emph{J. Phys.
  Chem. C} \textbf{2021}, \emph{125} (32), 17612\mynobreakdash \mynobreakdash
  17621. \url{doi: 10.1021/acs.jpcc.1c04239}.


\setboolean{nobreakdashused}{false}\bibitem[Abraham et~al.(2015)Abraham,
  Murtola, Schulz, P{\'{a}}ll, Smith, Hess, and Lindahl]{Abraham2015}
Abraham,~M.~J.; Murtola,~T.; Schulz,~R.; P{\'{a}}ll,~S.; Smith,~J.~C.;
  Hess,~B.; Lindahl,~E. \emph{SoftwareX} \textbf{2015}, \emph{1-2},
  19\mynobreakdash \mynobreakdash 25. \url{doi: 10.1016/j.softx.2015.06.001}.


\setboolean{nobreakdashused}{false}\bibitem[Plimpton(1995)]{Plimpton1995}
Plimpton,~S. \emph{J. Comput. Phys.} \textbf{1995}, \emph{117} (1),
  1\mynobreakdash \mynobreakdash 19. \url{doi: 10.1006/jcph.1995.1039}.


\setboolean{nobreakdashused}{false}\bibitem[Stuart et~al.(2000)Stuart, Tutein,
  and Harrison]{Stuart2013}
Stuart,~S.~J.; Tutein,~A.~B.; Harrison,~J.~A. \emph{J. Chem. Phys.}
  \textbf{2000}, \emph{112} (14), 6472\mynobreakdash \mynobreakdash 6486.
  \url{doi: 10.1063/1.481208}.


\setboolean{nobreakdashused}{false}\bibitem[Nos{\'{e}}(1984)]{Nose1984}
Nos{\'{e}},~S. \emph{J. Chem. Phys.} \textbf{1984}, \emph{81} (1),
  511\mynobreakdash \mynobreakdash 519. \url{doi: 10.1063/1.447334}.


\setboolean{nobreakdashused}{false}\bibitem[Hoover(1985)]{Hoover1985}
Hoover,~W.~G. \emph{Phys. Rev. A} \textbf{1985}, \emph{31} (3),
  1695\mynobreakdash \mynobreakdash 1697. \url{doi: 10.1103/PhysRevA.31.1695}.


\setboolean{nobreakdashused}{false}\bibitem[Han et~al.(1997)Han, Globus, Jaffe,
  and Deardorff]{Han1997}
Han,~J.; Globus,~A.; Jaffe,~R.; Deardorff,~G. \emph{Nanotechnology}
  \textbf{1997}, \emph{8} (3), 95\mynobreakdash \mynobreakdash 102. \url{doi:
  10.1088/0957-4484/8/3/001}.


\setboolean{nobreakdashused}{false}\bibitem[Robertson et~al.(1994)Robertson,
  Dunlap, Brenner, Mintmire, and White]{Robertson1994}
Robertson,~D.; Dunlap,~B.; Brenner,~D.; Mintmire,~J.; White,~C. \emph{MRS
  Proc.} \textbf{1994}, \emph{349} (Md), 283. \url{doi: 10.1557/PROC-349-283}.


\setboolean{nobreakdashused}{false}\bibitem[Volokitin and
  Persson(2017)]{Volokitin}
Volokitin,~A.~I.; Persson,~B.~N. \emph{{Electromagnetic Fluctuations at the
  Nanoscale}};
\newblock NanoScience and Technology;
\newblock Springer Berlin Heidelberg: Berlin, Heidelberg, 2017;
\newblock \url{doi: 10.1007/978-3-662-53474-8}.


\setboolean{nobreakdashused}{false}\bibitem[Chenoweth et~al.(2008)Chenoweth,
  van Duin, and Goddard]{Chenoweth2008}
Chenoweth,~K.; van Duin,~A. C.~T.; Goddard,~W.~A. \emph{J. Phys. Chem. A}
  \textbf{2008}, \emph{112} (5), 1040\mynobreakdash \mynobreakdash 1053.
  \url{doi: 10.1021/jp709896w}.


\setboolean{nobreakdashused}{false}\bibitem[Wang et~al.(2018)Wang, Zhang, Han,
  and E]{Wang2018}
Wang,~H.; Zhang,~L.; Han,~J.; E,~W. \emph{Comput. Phys. Commun.} \textbf{2018},
  \emph{228}, 178\mynobreakdash \mynobreakdash 184. \url{doi:
  10.1016/j.cpc.2018.03.016}.


\end{thebibliography}

\end{document}